# Unraveling structural, electronic, and magnetic ambiguities in $Pb_{1-\delta}CrO_3$ with an insulating charge-transfer band structure

Jian Chen[1,2], Guozhu Song[1,3], Han Ge[1], Antonio M. Dos Santos[4], Liusuo Wu[1,2], Yusheng Zhao[1,5], Shanmin Wang[1,2,*]

[1]*Department of Physics, Southern University of Science and Technology, Shenzhen, Guangdong, 518055, China*
[2]*Quantum Science Center of Guangdong-Hongkong-Macao Greater Bay Area, Shenzhen, Guangdong, 518045, China*
[3]*Department of Physics, City University of Hong Kong, Tat Chee Avenue, Kowloon, Hong Kong, 999077, China*
[4]*Spallation Neutron Source, Oak Ridge National Laboratory, Oak Ridge, Tennessee 37831, USA*
[5]*College of Science, Eastern Institute of Technology, Ningbo, Zhejiang, 315200, China*

**Abstract**

As a recently-identified Mott system, $PbCrO_3$ remains largely unexplored, especially for its band structure, leading to many contentious issues on its structural, electronic, and magnetic properties. Here we present a comprehensive study of two different $Pb_{1-\delta}CrO_3$ ($\delta = 0$ and 0.15) samples with involving atomic deficiency prepared under pressure. By means of the state-of-the-art diffraction techniques, crystal structure of $PbCrO_3$ is definitively determined to adopt the pristine $Pm\bar{3}m$ symmetry, rather than other previously misassigned structures of M2-$Pm\bar{3}m$ and $Pmnm$. The two materials exhibit a similar charge-transfer-type insulating band structure, and the charge-transfer effect splits both $Cr2p$ and $Pb4f$ orbitals, rationalizing doublet splitting of the associated spectral lines. Nearly identical nominal cationic valence states of $Cr^{4+}$ and $Pb^{2+}$ are identified for this oxide system, hence calling into question the validity of recently-proposed charge disproportionation mechanisms. Besides, $Pb_{0.85}CrO_3$ exhibits an anomalously higher Néel temperature of ~240 K than that of $PbCrO_3$ (i.e., ~200 K), likely due to the deficiency-induced enhancements of Cr: $3d$-O:$2p$ orbital overlap and magnetic exchange. These findings provide many solid evidences to look into the fundamental properties of this important material system.

**Introduction**

The 3*d* transition-metal (TM) oxide perovskites ABO$_3$ (A = main group or rare-earth metals and B = 3*d* TMs) have attracted considerable attention over past decades [1-4], due to their intriguing structural and magnetic properties that primarily originate from both the strongly correlated 3*d* electrons and competition of covalent A-O and B-O hybridizations. These correlation and competitions are often sensitive to external fields such as pressure (P) and temperature (T) [5-8], making their properties largely tunable, which is important for study of many exotic phenomena in condensed-matter physics.

Among them, PbCrO$_3$ represents one of the most exciting materials and can only be synthesized at pressure above 6 GPa, adopting a cubic structure with an antiferromagnetic (AFM) insulating ground state at ambient pressure [9-11]. This oxide can readily be doped by Ti or V atoms to form PbCr$_{1-x}$M$_x$O$_3$ (M = Ti or V) with a wide range of adjustable doping concentration, implying excellent turnability of its properties [12,13]. Very recently, PbCrO$_3$ has been identified to be a unique Mott system [14], showing a first-order isostructural transition, coupled with an insulator-metal transition [15-17]. Interestingly, this material also exhibits critical behaviors under P-T conditions with an critical endpoint at 430 K and 4.82 GPa, beyond which the second-order transition starts to occur; near the Mott criticality, the lattice deforms anomalously, characterized by giant viscoelasticity that likely results from the coupling of lattice elasticity and electron viscosity [14]. These important results would be pivotal for unveiling the long-standing mysteries underlying the Mott transition.

However, there still exist a number of controversial reports on its crystal structure and electronic and magnetic properties, which have severely impeded exploring the associated phenomena as observed in this material system.

In fact, PbCrO$_3$ can only be prepared at pressure and subjected to an isostructural transition with a ~7% volume expansion during decompression, so that the high-density crystalline imperfections would be inevitably introduced into the recovered crystals with plentiful lattice distortions, as manifested by anomalous x-ray diffraction peak broadening (refs. [11,15,17,18]) and lattice-strain-modulated superlattices (ref. [19]). It is noted that previous spectral measurements have observed doublet splitting of the core-level Cr: $L_{2,3}$ and Pb: 4$f$ spectral lines [16,18,20,21], which, however, have been tentatively interpreted by charge disproportionation mechanisms, giving rise to two differently configured phases of PbCr$^{3+}_{2/3}$Cr$^{6+}_{1/3}$O$_3$ and Pb$^{2+}_{0.5}$Pb$^{4+}_{0.5}$CrO$_3$ [18,21]. Based on those experimental reports, a couple of structures have been assigned for PbCrO$_3$, including cubic M2-$Pm\bar{3}m$ and orthorhombic $Pmnm$ models [18,21]. Although the earlier proposed "CrPbO$_3$" model (refs. [15,22]) has been definitively ruled out by our neutron diffraction experiment [17], it is still occasionally mentioned [20]. More conflicting matter further is that two apparently different Néel temperatures (T$_N$) of 240 K (refs. [9,23]) and 200 K (ref. [17]) have been reported for PbCrO$_3$, which may be related to the atomic deficiency. However, to date, such contentious issues of this material system have not been well clarified as of yet.

Remarkably, those contradictory reports are presumably due to the inadequate understanding of band structure of this oxide. Judged from its nominal valence state of $Cr^{4+}$, PbCrO$_3$ should has a charge-transfer-type band structure rather than a Mott-Hubbard-type one (Fig. 1), because the involved on-site Coulomb repulsion (i.e., U ≈ 5.5 eV) is greater than the charge-transfer energy (i.e., Δ ≈ 3 eV) [24], similar to that of V$_2$O$_3$ and (La$_{1-x}$Sr$_x$)$_2$CuO$_{4-y}$ with U > Δ [25,26]. In this case, the ligand O: 2$p$ orbital is located above the lower Cr: 3$d$ orbital of the valence band, and the overlap of them can form the covalent Cr: 3$d$–O: 2$p$ hybridization. As a result, the O: 2$p$ → Cr: 3$d$ charge transfer can be easily induced during various spectral processes, producing the ligand hole and 3$d$-like electrons, hence the multiple final states for splitting the core-level spectral lines, which can be arguably described by consideration of the configuration interactions and ligand-hole cluster theory [24,27-29].

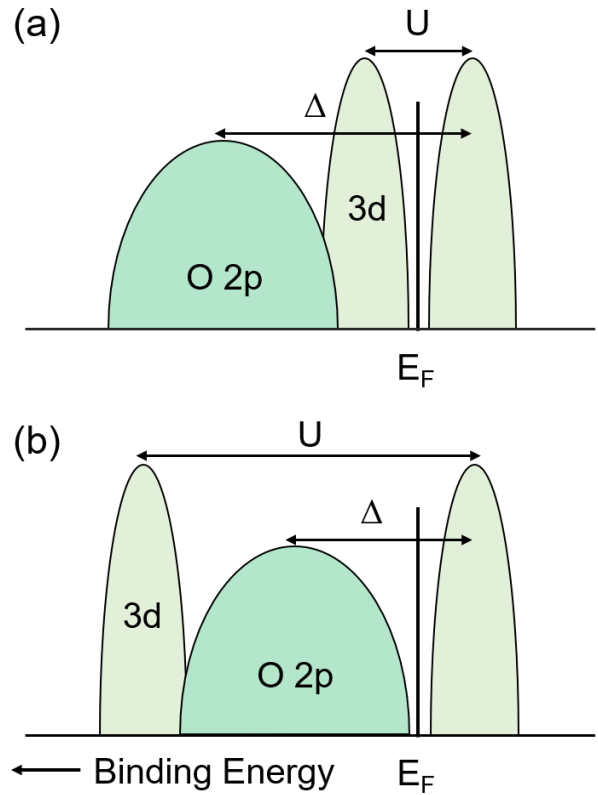

**Fig. 1. Schematics of band structures of two different 3$d$ insulators. (a)** Mott-Hubbard insulators with U < Δ. **(b)** Charge-transfer insulators with U > Δ. U and Δ represent the on-site Coulomb repulsion and charge transfer energy, respectively, referring to refs. [30-32] for more information.

Bearing these issues in mind, we perform high-P synthesis of this material system, leading to successful attainment of a substoichiometric compound of $Pb_{0.85}CrO_3$, in addition to $PbCrO_3$, with Pb atomic deficiency. The crystal structures of samples are definitively resolved to be the pristine $Pm\bar{3}m$ symmetry with a similar insulating charge-transfer-type band structure. The doublet splitting of core-level Cr2$p$ and Pb4$f$ spectral peaks is also observed, which is excellently interpreted by the charge-transfer effect. Besides, other ambiguities regarding magnetic and electronic properties of the material system are also elucidated in details.

**Experimental details**

***Synthesis.*** Commercially available $PbCrO_4$ (> 99.7%) was purchased and used as a starting material for the synthesis of $Pb_{1-\delta}CrO_3$ ($\delta$ = 0 and 0.15) samples by $O_2$ degassing under high P-T conditions. High-P experiments were carried out in a DS 6 × 10 MN cubic press installed in the high-P Lab of SUSTech with well calibrated pressure and temperature [33]. Before the experiment, $PbCrO_4$ starting powders were compacted into a cylindrical pellet of 8.3 mm in diameter and 5 mm in height and then encapsulated by a NaCl capsule to avoid possible contamination because of the highly reactive activity of $PbCrO_4$, followed by being assembled with pre-prepared cell parts. In each experimental run, we first compressed the sample to a target pressure of ~6 GPa and subsequently heated it to 1100 ºC with a heating duration of 90 min before quenching by directly turning off the heating power. The recovered sample was initially washed with water to remove NaCl capsule material and then treated with highly concentrated

NaOH aqueous solutions to dissolve unreacted PbCrO$_4$. After multiple washes of such-treated sample with distilled water, nearly impurity-free sample was obtained. Using a needle by hand, crystals of two different morphologies were readily separated under a microscope, corresponding to two phases of Pb$_{1-\delta}$CrO$_3$.

***Characterizations.*** The final products were checked by x-ray diffraction (XRD, Rigaku SmartLab diffractometer) with a Cu Kα target to determine the crystal structure. The samples' morphologies and compositions and valence states of the involved elements were measured by scanning electron microscopy (SEM, FEI Noval NanoSEM 450), energy-dispersive x-ray spectroscopy (EDS), and x-ray photoelectron spectroscopy (XPS, PHI 5000 Versaprobe III), respectively. Based on single-crystal samples, low-T electrical resistivity measurements were performed by a four-probe method, using a PPMS (Quantum Design). Low-T magnetic susceptibilities of samples were measured at an external field of 1000 Oe, using an MPMS3 SQUID magnetometer (Quantum Design); besides, the samples' magnetization was determined by sweeping the magnetic field between −7 to 7 T at a set of desired temperatures of 2, 50, 100, 150, 200, and 300 K, respectively. For each of the two as-obtained Pb$_{1-\delta}$CrO$_3$ phase, the magnetic measurement was conducted based on a number of separated single crystals of phase-pure Pb$_{1-\delta}$CrO$_3$. Time-of-flight neutron powder diffraction (NPD) experiment of PbCrO$_3$ was performed using the SNAP beamline at the Spallation Neutron Source (SNS) of ORNL, USA, and the NPD data was taken at ambient conditions by two banks of detectors (i.e., Bank1 and Bank2) with different d-spacing coverages of 0.8 – 3.2 Å and 1.8 – 5.0 Å, respectively.

**Results and discussion**

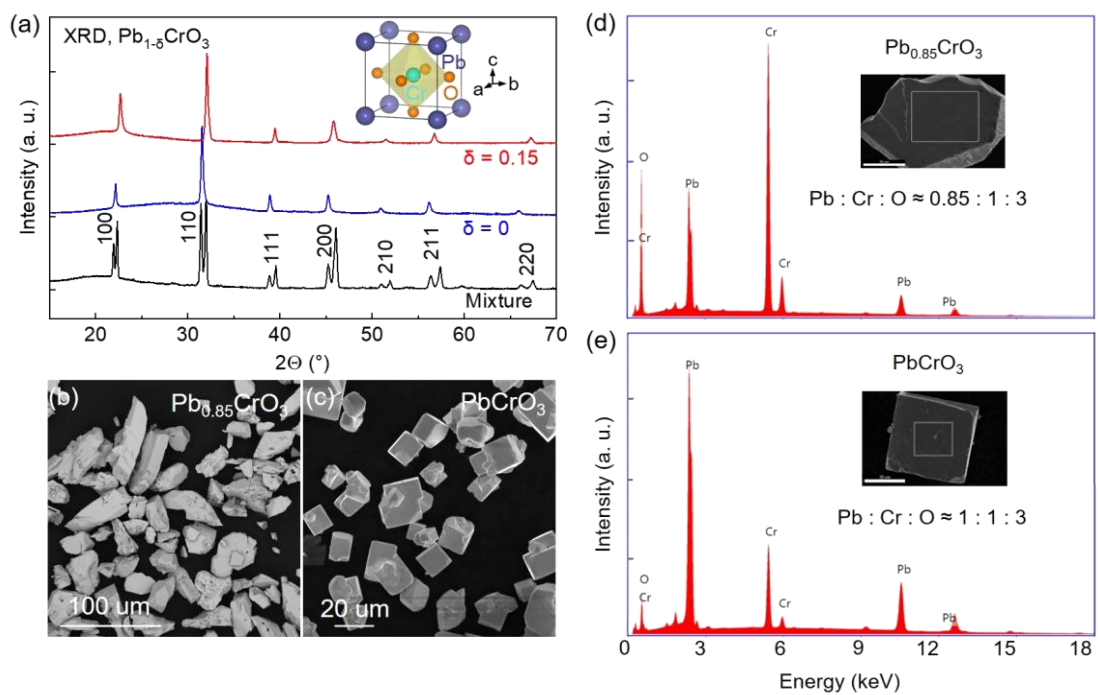

**Fig. 2. Crystal structures, morphologies, and compositions of as-prepared $Pb_{1-\delta}CrO_3$. (a)** XRD patterns taken at ambient conditions. Both $Pb_{0.85}CrO_3$ and $PbCrO_3$ coexist in the recovered sample, but they can be separated in terms of their crystal morphologies. **(b)-(c)** SEM images of the separated $Pb_{0.85}CrO_3$ and $PbCrO_3$ crystals with irregular and regular shapes, respectively. **(d)-(e)** EDS spectra. Insert in each panel is a selected crystal with the framed areas for EDS measurements.

The final product shows a set of sharp XRD peaks with a clean background (Fig. 2(a)), signaling excellent crystallinity of the recovered sample. However, each reflection line seems split into a doublet; careful analysis indicates that two phases are involved and can be indexed by the same $Pm\bar{3}m$ symmetry with different lattice parameters of $a$ = 4.0112(2) and 3.9813(5) Å, respectively. The former is close to that of cubic $PbCrO_3$ (i.e., ~4.01 Å) as previously reported [9,15], while the latter is ~1% smaller than that of the former mainly due to the atomic deficiency of Pb, giving rise to an off-stoichiometric $Pb_{1-\delta}CrO_3$ as will be discussed below. Indeed, there are two distinct crystal morphologies of the sample can be clearly discerned, displaying regular cuboid and irregular shapes, respectively, with crystallite sizes of ~20 μm and 50 – 100 μm (Fig. 2(b)-2(c) and Fig. S1). Such differently-shaped crystals can be readily separated under a microscope, and each of the thus-separated crystals is phase-pure and gives a single-phase XRD pattern without peak splitting (Fig. 2(a)).

According to elemental analysis (Fig. 2(c)-2(d)), the two phases are compositionally constituted by Pb, Cr, and O with atomic ratios of ~1 : 1 : 3 and ~0.85 : 1 : 3 for regular and irregular crystals, respectively, in excellent agreement with the refined Pb atomic occupancies of 0.85(2) and 1, based on the Rietveld analysis of XRD data (Fig. S2). Clearly, these results indicate the presence of Pb deficiency in the crystals with irregular shapes, while the regular cuboid crystals are nearly stoichiometric $PbCrO_3$. The thus-determined Pb deficiency concentrations in the two $Pb_{1-\delta}CrO_3$ samples are δ = 0.15(2) and 0, respectively, giving actual compositions of $Pb_{0.85}CrO_3$ and $PbCrO_3$.

To address the structural ambiguities of $PbCrO_3$, combined with simulations, we conduct careful analysis of our experimental XRD and NPD data in Fig. 3, based on a couple of the already-reported models (Fig. 3(a)). The observed XRD peaks of $PbCrO_3$ are asymmetrically broadened toward the high-2Θ values (Fig. 3(b)), which is consistent with previous reports [11,19], reflecting the existence of a microscopically compressive stress state in the crystals [34,35]. We note that samples were synthesized at high pressure of above 6 GPa and underwent an isostructural transition with a ~7% volume expansion during decompression [14,15,17]. Such a large volume change during transition would lead to the nucleation of many nuclei of low-P new phase in the matrix of high-P parent phase [36]. Thus, plentiful crystalline imperfections should be produced including atomic defects and lattice distortions with large lattice stress. Some of them are eventually preserved in the final crystals, accounting for both the observed anomalous XRD peak broadening and complexity of microstructures such as microdomain textures and atomic-deficiency modulated supercells [19]. According to those experimental observations, the afore-mentioned M2-$Pm\bar{3}m$[18] and orthorhombic *Pmnm* [21] have thus been proposed recently for $PbCrO_3$ (Fig. 3a), involving charge disproportionation of either Cr or Pb. Besides, a $Pm\bar{3}m$-"$CrPbO_3$" model with anti-sites has also been initially adopted to explain the pressure-induced isostructural transition [15]. As a result, the actual crystal structure of $PbCrO_3$ is still under debate.

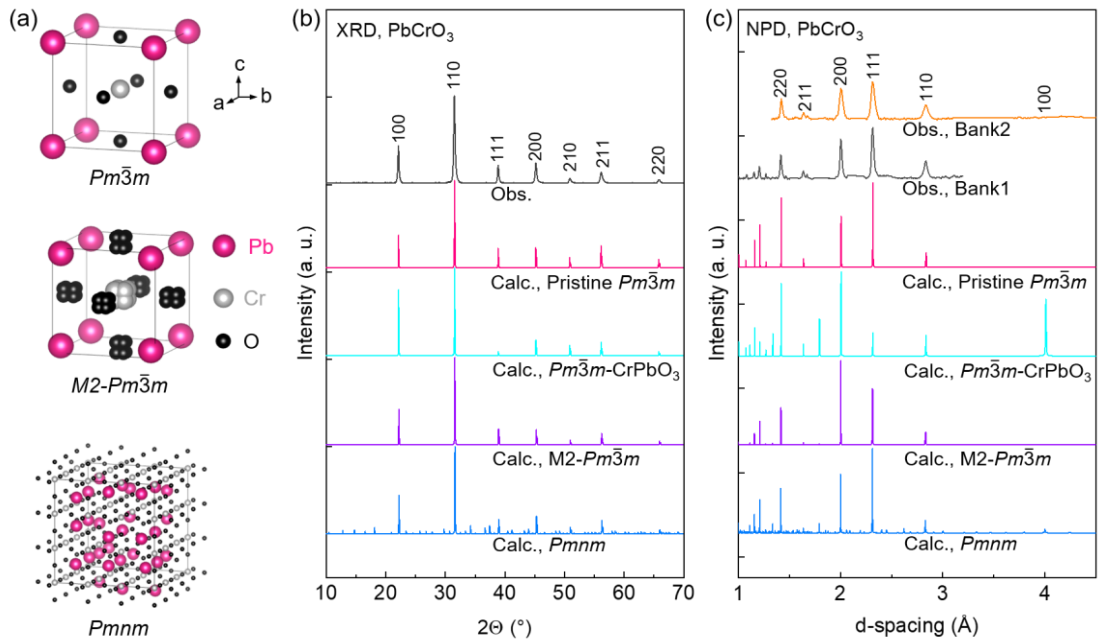

**Fig. 3. XRD and NPD patterns of stoichiometric $PbCrO_3$.** (**a**) Three different models of crystal structures including the pristine $Pm\bar{3}m$, M2-$Pm\bar{3}m$ [18], and *Pmnm* [21]. (**b**) XRD patterns. (**c**) NPD patterns. The simulated XRD and NPD patterns of three different models are added for comparison. The simulations are also performed using the anti-site "$CrPbO_3$" model with an atomic position interchange between Pb and Cr of its pristine $Pm\bar{3}m$-$PbCrO_3$, as proposed earlier [15].

By comparison of the simulated XRD and NPD patterns with the experimental ones (Fig. 3(b)-3(c)), the *Pmnm* can be quickly excluded because it gives many extra reflection lines that do not exist in the experiment. For the M2-$Pm\bar{3}m$ model with both Pb and Cr atoms located at the general Wyckoff sites of 8g (*x, x, x*), there is a large intensity mismatching of the simulated and measured NPD patterns, especially for the 200 and 111 peaks (Fig. 3(c)), although the simulated XRD pattern is almost indistinguishable with that of the pristine $Pm\bar{3}m$. The anti-site "CrPbO$_3$" model can also be ruled out, as the 100 peak of NPD pattern is systematically absent. This is because the coherent neutron scattering lengths of $f_{Pb}$, $f_{Cr}$, and $f_O$ coincidentally stratify the relation $f_{Pb} \approx f_{Cr} + f_O$ [37], giving rise to a nearly zero structure factor, $(F_{100})^2$, for neutron diffraction (i.e., $(F_{100})^2 \approx 0$) [17]. In contrast, by using the pristine $Pm\bar{3}m$ model, both the XRD and NPD patterns can be excellently reproduced, indicating it is the most suitable structure for this material system.

**Table 1. Summary of refined lattice parameters for Pb$_{0.85}$CrO$_3$ and PbCrO$_3$, based on the Rietveld analysis of XRD data taken at ambient conditions.**

|  | Pb$_{0.85}$CrO$_3$ | PbCrO$_3$ |
|---|---|---|
| Symmetry | Cubic, $Pm\bar{3}m$ (No. 221) | Cubic, $Pm\bar{3}m$ (No. 221) |
| Cell parameter (Å) | 3.9813(5) | 4.0112(2) |
| Cell volume (Å$^3$) | 63.107(1) | 64.537(3) |
| Density (g/cm$^3$) | 7.307 | 7.891 |
| Wyckoff site | Pb: 1a (0, 0, 0) | Pb: 1a (0, 0, 0) |
|  | Cr: 1b (1/2, 1/2, 1/2) | Cr: 1b (1/2, 1/2, 1/2) |
|  | O: 3c (0, 1/2, 1/2) | O: 3c (0, 1/2, 1/2) |
| Occupancy (Pb) | 0.85(2) | 1 |
| $d_{Pb-O}$ (Å) | 2.8152(3) | 2.8363(2) |
| $d_{Cr-O}$ (Å) | 1.9907(3) | 2.0056(2) |
| wRp (%), $\chi^2$ | 9.51, 1.78 | 7.33, 1.51 |

The thus-refined crystal structures for both PbCrO$_3$ and Pb$_{0.85}$CrO$_3$ are listed in Table 1. Clearly, the Cr atoms are octahedrally coordinated by six O atoms (Fig. 2(a)), while the Pb atoms are bonded by twelve O atoms to form tetradecahedral coordinates. It is noted that Pb$_{0.85}$CrO$_3$ can only be obtained in a narrow temperature range of 1050 - 1100 ºC at 6 GPa with a prolonged heating duration over 90 min. Attempts to prepare other Pb$_{1-\delta}$CrO$_3$ with different δ values are unsuccessful, indicating a unique delicate re-balance of the stressed Pb-O and Cr-O bonds for stabilizing its crystal structure, as will be mentioned later.

The decrease of lattice parameter in Pb$_{0.85}$CrO$_3$ can be rationalized from the presence of atomic deficiency. In fact, there exists a strong competition between covalent Cr: $t_{2g}$–O: $p_\pi$ and Pb: $sp^3$–O: $p_\pi$ bonding states [1]. Thus, the Cr-O bonding would be enhanced if the competition from Pb is relaxed by introducing Pb deficiency. As a consequence, the Cr-O bond is shortened to be 1.9907 Å from its pristine length of 2.0056 Å with a 0.74% decrease (see Table 1), which in turn alters the stress state of Pb-O bonds and subsequently varies the Cr valence state, eventually leading to different magnetic, electronic, and structural properties of this oxide.

Low-T resistivities of both samples are measured in Fig. 4 to study their electrical transport behaviors. With decreasing temperature, the two samples display a similarly rapid increase of resistivity by more than three orders of magnitude, a characteristic of insulators, consistent with previous reports of PbCrO$_3$ [17,18]. Obviously, Pb$_{0.85}$CrO$_3$ has a relatively low resistivity, nearly one order of magnitude smaller than that of PbCrO$_3$ at 300 K. As mentioned above, due to the competition of Pb-O and Cr-O bonding states the atomic deficiency in $Pb_{0.85}CrO_3$ can promote a substantial overlap of Cr: $3d$ – O: $2p$ orbitals for narrowing the bandgap, as manifested by largely reduced resistivity. In addition, a relatively low concentration of crystalline imperfections should be involved in $Pb_{0.85}CrO_3$, because it is more mechanically robust than that of the stoichiometric sample as a result of the reduced defects. A lower concentration of defects is really favorable for improving the sample's conductivity.

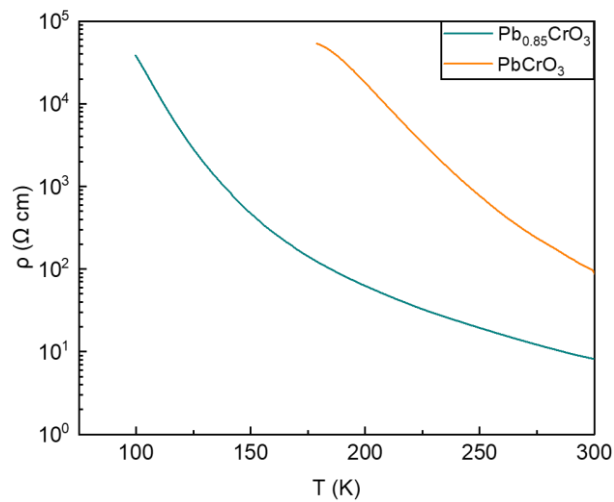

**Fig. 4. Determined low-T electrical resistivities for both $Pb_{0.85}CrO_3$ and $PbCrO_3$, based on a four-probe method.**

To gain the electronic properties of both materials, we perform the core-level XPS measurements with a focus on Cr$2p$ and Pb$4f$ states, as presented in Fig. 5(a)-5(b). Due to the spin-orbit coupling, the associated XPS spectra are split into two main components of $2p_{1/2}$ and $2p_{3/2}$ for Cr and $4f_{5/2}$ and $4f_{7/2}$ for Pb, respectively. On a closer inspection, each main component line of Cr$2p$ and Pb$4f$ is actually a doublet, which can be explained as a result of charge-transfer-induced multiple final states by considering the configuration interaction [24,27]. Because of the typical {CrO$_6$} octahedral coordination in $Pb_{1-\delta}CrO_3$, the five Cr: $3d$ orbitals are split into two groups of degenerate orbitals of $e_g$ ($d_{z^2}$ and $d_{x^2-y^2}$) and three $t_{2g}$ ($d_{xy}$, $d_{yz}$, and $d_{zx}$) states by the crystalline fields (see Fig. 5(c)) [38]. Such $e_g$ orbitals are directed toward near-neighbor anions to form a strong Cr: $e_g$–O: $2p$ hybridization, which plays an important role for producing the charge transfer effect. In this regard, a certain fraction of the otherwise localized electrons in O: $2p$ orbitals can be transferred back to Cr: $3d$ orbitals, resulting in two different major electronic states of Cr: $3d^2$ and $3d^3\underline{L}$ (where $\underline{L}$ denotes a ligand hole in the O: $2p$ orbital), on the basis of the metal-ligand cluster theory.[27,39]

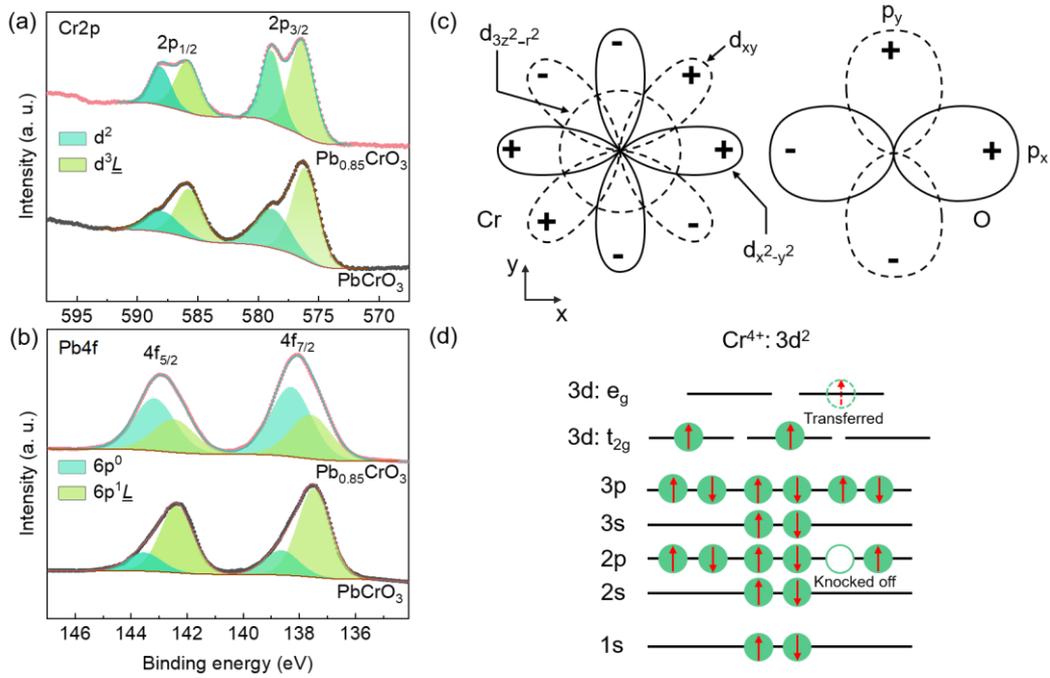

**Fig. 5. XPS spectra of Pb$_{0.85}$CrO$_3$ and PbCrO$_3$ taken at ambient conditions based on single-crystal samples. (a)** Cr2$p$ lines. **(b)** Pb4$f$ lines. **(c)** Configuration of Cr: $3d$ and O: $2p$ orbitals. **(d)** Schematic of an x-ray photoemission process for the core-level Cr2$p$ orbital with a charge transfer effect.

In the process of the Cr2$p$ photoemission, the incident high-energy x-rays first knock off one of Cr: $2p^6$ electrons for the formation of $2p^5$ states (Fig. 5(d)); the interactions of Cr: $3d^2$ and $3d^3\underline{L}$ with the thus-formed $2p^5$ give rise to two major final states of $2p^53d^2$ and $2p^53d^3\underline{L}$, rationalizing the observed doublet splitting of Cr2$p_{1/2}$ and $2p_{3/2}$ orbitals in both $Pb_{0.85}CrO_3$ and $PbCrO_3$. Note that the valance states of Cr in the two samples are nearly similar with a nominal value of +4. The term of $\Delta$ is thus estimated to be ~3.3 eV and much smaller than that of $U \approx 5.5$ eV [24], giving a charge-transfer-type band structure with the O: $2p$ band located above the lower Mott-Hubbard $3d$ band (Fig. 1). This indicates that the charge transfer between the Cr: $e_g$ and O: $2p$ orbitals is energetically more favored, rather than between the nearest neighboring Cr: $t_{2g}$–Cr: $t_{2g}$ orbitals. Therefore, the main peaks of the Cr: $2p$ doublet in Fig. 5(a) should correspond to the $2p^5 3d^3 \underline{L}$ state, while the satellite peak arises from the $2p^5 3d^2$ state (Fig. 5(a)). For $Pb_{0.85}CrO_3$, the intensity ratio of the main to satellite peak is relatively smaller, due to the deficiency-enhanced Cr-O covalency having a decreased charge transfer [24,27]. A similar Cr: $2p$ doublet also occurs in $CrO_2$ with an enhanced intensity of the main peak relative to that of $PbCrO_3$ (Fig. S3(a)), which suggests a smaller Cr: $3d$–O: $2p$ overlap in $PbCrO_3$ for reducing the charge transfer effect.

Besides, an obvious peak sharpening can also be observed in $Pb_{0.85}CrO_3$ (Fig. 5(a)), because the multiplet splitting of both the $3d^2$ and $3d^3 \underline{L}$ states are relatively smaller than that of $PbCrO_3$. Such a subtle difference in peak splitting should be closely attributed to the intra-atomic Coulomb and exchange interactions, based on previous theoretical calculations [27,40]. The sharpest Cr: $2p$ lines can be seen in $Na_2CrO_4$ (i.e., $Cr^{6+}$) with an empty $d$ orbital (Fig. S3(a)); in this case the intra-atomic interactions are negligible, causing a nearly zero multiplet splitting. For $CrCl_3$ and $Cr_2O_3$ with a nominal $Cr^{3+}$ valence state, they belong to the Mott-Hubbard insulators with $\Delta > U$ [24], without the presence of peak doublets.

A similar splitting of the Pb4*f* orbitals is occurred and can also be interpreted by the charge transfer effect, involving two major final states of $6p^0$ and $6p^1\underline{L}$. The occurrence of charge transfer in Pb-bearing compounds is not unexpected, as Pb has various valence states in analogy to transition metals. According to the band structure calculations of high-P PbCrO$_3$ [18] and its sister material of PbTiO$_3$ [41], the crystalline field splits the Pb: 6*p* orbitals into the bonding and antibonding states, which is different from that of the Cr: 3*d* band that is split by the on-site Coulomb repulsion. The resultant Pb: 6*p* bonding states are largely overlapped with the O: 2*p* band to form a covalent Pb: 6*p*–O: 2*p* hybridization. Therefore, the main and satellite peaks of Pb4$f_{5/2}$ and 4$f_{7/2}$ correspond to the $6p^1\underline{L}$ and $6p^0$ states, respectively. By introducing the Pb deficiency, the probability for charge transfer from O: 2*p* to Pb: 6*p* is accordingly lowered, which should result in a relative intensity decrease of the main peaks, in excellent consistence with the observed in Pb$_{0.85}$CrO$_3$ (Fig. 5(b)). As expected, PbTiO$_3$ shows a similar Pb4*f* peak splitting (Fig. S3(b)), but the energy separation between the main and satellite lines (i.e., 0.95 eV) are smaller than that of PbCrO$_3$ (i.e., 1.16 eV). Nearly pure ionicity of Pb$^{2+}$ is involved in PbCrO$_4$ and PbX$_2$ (X = F, Cl, and Br) and gives rise to singlet peaks, likely because the relevant charge transfer is difficult without a suitable Pb: 6*p*–O: 2*p* orbital overlap (Fig. S3(b)). The situations of both PbO$_2$ and PbO with certain Pb-O hybridizations are completely changed (Fig. S3(b)), showing a clear doublet peak splitting.

Obviously, all these evidences strongly suggest that both PbCrO$_3$ and Pb$_{0.85}$CrO$_3$ are charge-transfer-type insulators. The peak doublet splitting of Cr2*p* and Pb4*f* is primarily related to the charge transfer effect, which can also be applicable for elucidating many other complex spectral processes such as x-ray emission and absorption [27,42,43]. Besides, the nominal cationic valence states of $Pb_{1-\delta}CrO_3$ are clearly determined to be close to $Cr^{4+}$ and $Pb^{2+}$, respectively, hence invalidating previously reported mechanisms of charge disproportionation that were proposed based on inappropriate data analysis of spectral line splitting without considering the charge transfer effect [16,18,20,21].

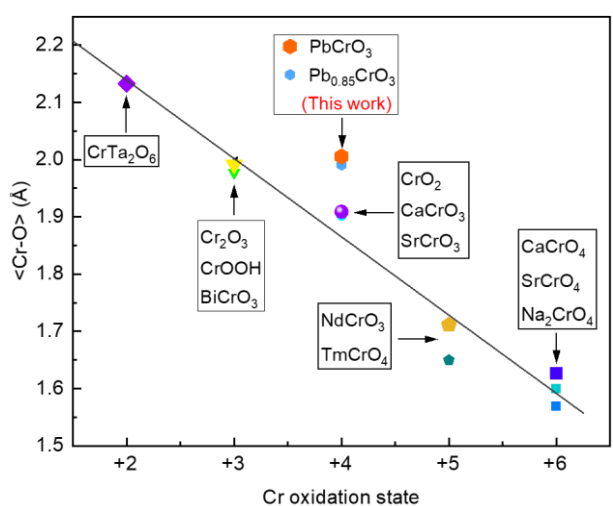

**Fig. 6. Plot of Cr-O bond lengths vs. oxidation state of Cr, based on a series of previously reported data of Cr-bearing oxides** [44]**.** The black line denotes a linear regression of the data.

As shown in Fig. 6, both $PbCrO_3$ and $Pb_{0.85}CrO_3$ show a similar anomalously large Cr-O bond length of ~2 Å, more than 5% longer than that of other oxides with the same nominal valence state of $Cr^{4+}$, making it clearly deviated from the linear variation of the Cr-O bond length against the Cr oxidation state. This phenomenon is closely related to the mismatching of Pb-O and Cr-O bond lengths. In fact, by summation over the associated ionic radii [45], the equilibrium bond lengths of Pb-O and Cr-O can be easily calculated to be 2.89 and 1.95 Å, respectively, inconsistent with the x-ray measured values of ~2.83 and 2 Å (see Table 1). It is evident that the Pb-O is compressed, whereas the Cr-O is stretched. Such mismatching of bond lengths in this system would induce large stress that may be important for stabilizing its crystal structure and driving an interesting isostructural transition at pressure.

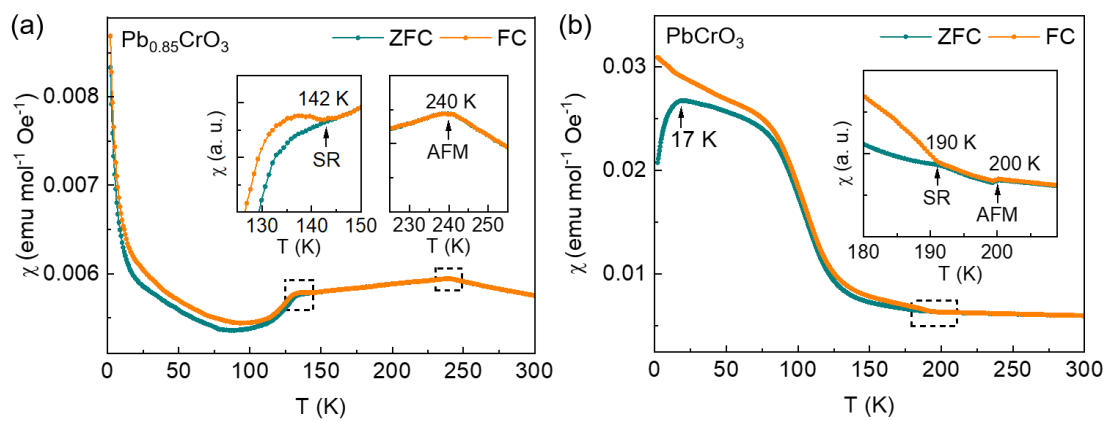

**Fig. 7. Magnetic measurements of $Pb_{0.85}CrO_3$ and $PbCrO_3$.** (a)-(b) Temperature dependence of the molar D.C. magnetic susceptibilities measured in an external field of 1000 Oe. Insert in each panel is an enlarged portion to show details. Arrows denote the magnetic transitions including antiferromagnetic (AFM) and spin re-orientation (SR) of magnetic domains.

As plotted in Fig. 7, the two materials show remarkably different magnetic properties. Apparently, the antiferromagnetic transition in $PbCrO_3$ occurs at $T_N \approx 200$ K, consistent with our previous neutron diffraction experiment [17]. However, a greater $T_N$ of 240 K is identified for $Pb_{0.85}CrO_3$, reflecting a large enhancement of magnetic exchange due to the shortened Cr-O bond length with a increased Cr: $3d$–O: $2p$ orbital overlap. Also noted is that previous studies reported a similar $T_N \approx 240$ K in this material system, but the involved samples were misassigned as "$PbCrO_3$" [9,23]. According to our present study, the real composition of their samples should actually be $Pb_{0.85}CrO_3$, rather than $PbCrO_3$. Below the $T_N$, the splitting of susceptibility curves, χ(T), of zero-field cooling (ZFC) and field cooling (FC) processes is occurred at 142 K for $Pb_{0.85}CrO_3$ and 190 K for PbCrO$_3$, respectively. Such splitting is suppressed by increase of the external field above 0.5 T (Fig. S4), indicating that the associated transitions may correspond to spin re-orientations (SR) of magnetic domains, as the magnetic polarization of those domains can be readily re-oriented under the external field.

An additional spin ordering in PbCrO$_3$ looks observable at 17 K, which is absent in Pb$_{0.85}$CrO$_3$. By analysis of these magnetic data using the Curie-Weiss method, the magnetic moments of both materials are determined to be 1.97 $\mu_B$ for Pb$_{0.85}$CrO$_3$ and 2.4 $\mu_B$ for PbCrO$_3$ (Fig. S4 and Table S3), respectively, close to the reported values by neutron diffraction measurements (i.e., 1.9 $\mu_B$ and 2.2 $\mu_B$) [9,17]. Besides, a weak ferromagnetic response is also observed in PbCrO$_3$ with profoundly enhanced magnetization, especially below 150 K, probably due to the grain size and surface effect that often result in a high proportion of uncompensated spins in the surfaces of sample grains [46-48]. This also indicates that more crystalline imperfections are involved in stoichiometric PbCrO$_3$ sample, probably because an anomalously larger volume change is involved at isostructural transition during decompression from its synthesis pressure.

**Conclusions**

In summary, by using a high-P decomposition method, we have successfully synthesized high-quality Pb$_{1-\delta}$CrO$_3$ ($\delta$ = 0 and 0.15) samples, showing a unique Pb atomic deficiency. Both materials adopt the same ordinary $Pm\bar{3}m$ symmetry with a nearly similar charge-transfer-type band structure. The doublet splitting of Cr2$p$ and Pb4$f$ XPS spectral lines is explained by the charge transfer effect, rather than the previously-proposed charge disproportionation, giving rise to nominal cationic valence states of $Cr^{4+}$ and $Pb^{2+}$. By introducing the Pb atomic deficiency, the Néel temperature is revealed to be profoundly promoted from ~200 K in $PbCrO_3$ to ~240 K in $Pb_{0.85}CrO_3$, mainly due to an increase of the Cr: $3d$–O: $2p$ orbital overlap with an enhanced magnetic exchange. In addition, many previous ambiguities regarding crystal and electronic structures and magnetic properties of this material system have also been well elucidated, which provide a solid foundation for exploring the associated Mott transition.

**Acknowledgments**

We acknowledge the financial support from the National Natural Science Foundation of China under Grant Nos. 12174175 and 12474013, the Guangdong Basic and Applied Basic Research Foundation under Grant No. 2022B1515120014, and the Shenzhen Basic Research Funds under Grants No. 20220815101116001 and No. JCYJ20220530113016038. The use of the SNAP diffractometer at the Spallation Neutron Source, a DOE Office of Science User Facility, was operated by the Oak Ridge National Laboratory, USA. Some experiments were supported by the Synergic Extreme Condition User Facility.

**Competing interests**

The authors declare no competing interests.

**Author Information**

**\*E-mail:** wangsm@sustech.edu.cn (S. Wang).

# Supplemental Material

# Unraveling structural, electronic, and magnetic ambiguities in $Pb_{1-\delta}CrO_3$ with an insulating charge-transfer band structure

Jian Chen[1,2], Guozhu Song[1,3], Han Ge[1], Antonio M. Dos Santos[4], Liusuo Wu[1,2], Yusheng Zhao[1,5], Shanmin Wang[1,2,*]

[1]*Department of Physics, Southern University of Science and Technology, Shenzhen, Guangdong, 518055, China*
[2]*Quantum Science Center of Guangdong-Hongkong-Macao Greater Bay Area, Shenzhen, Guangdong, 518045, China*
[3]*Department of Physics, City University of Hong Kong, Tat Chee Avenue, Kowloon, Hong Kong, 999077, China*
[4]*Spallation Neutron Source, Oak Ridge National Laboratory, Oak Ridge, Tennessee 37831, USA*
[5]*College of Science, Eastern Institute of Technology, Ningbo, Zhejiang, 315200, China*

E-mail: wangsm@sustech.edu.cn (S. Wang)

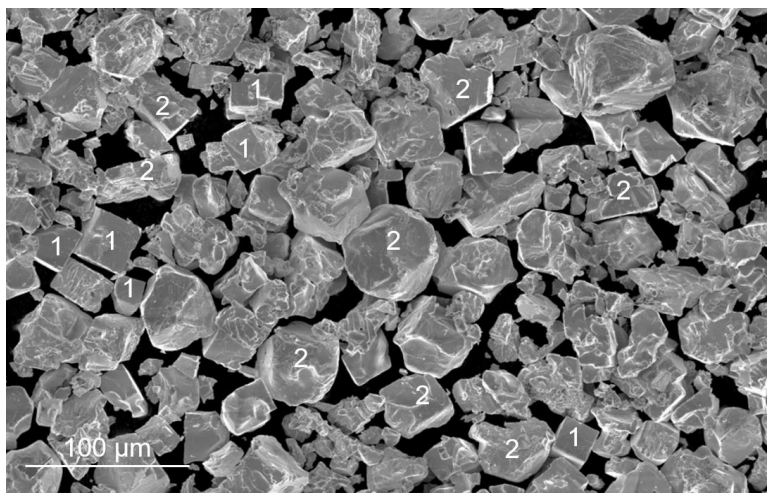

**Fig. S1. SEM images of the recovered sample containing both Pb$_{0.85}$CrO$_3$ and PbCrO$_3$** with two different crystalline morphologies of irregular and irregular shapes as denoted by "1" and "2", respectively.

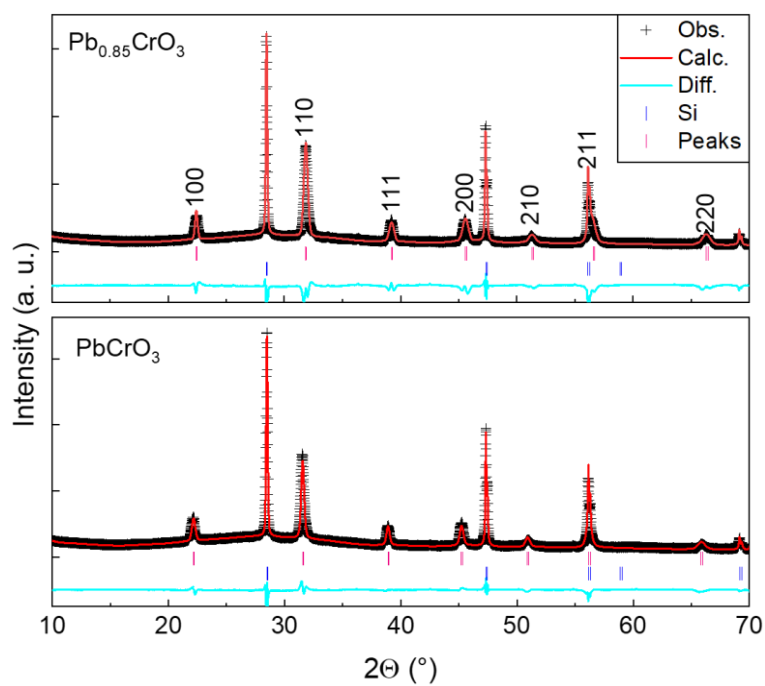

**Fig. S2. Refined XRD patterns for Pb$_{0.85}$CrO$_3$ and PbCrO$_3$.** The XRD data are collected at ambient conditions with a copper target (i.e., λ = 1.5406 Å). The Si powders are employed as an internal standard. The refined structural parameters are summarized in Table 1 of the main text.

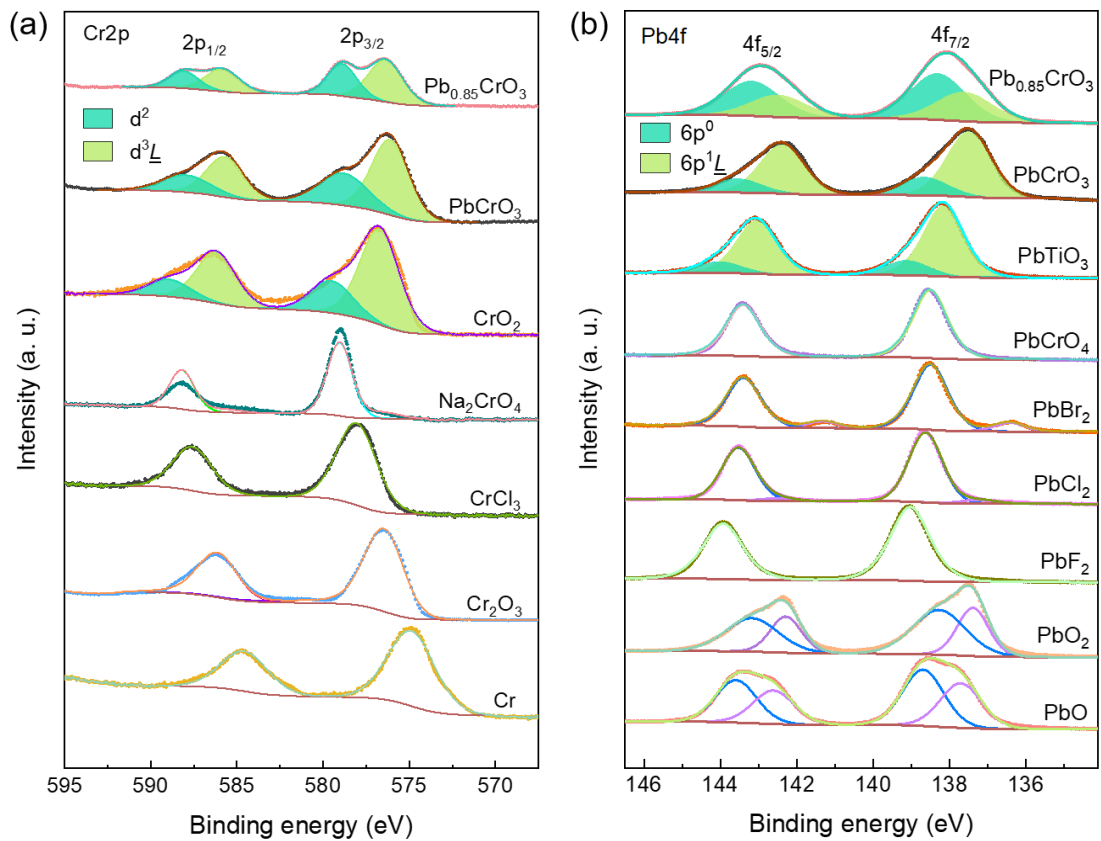

**Fig. S3. XPS spectra of the Cr2*p* and Pb4*f* based single-crystal samples of Pb₀.₈₅CrO₃ and Pb₀.₈₅CrO₃.** **(a)** Cr2*p* lines. **(b)** Pb4*f* lines. Experiments on other Cr-bearing compounds (i.e., $Na_2CrO_4$, $CrO_2$, $CrCl_3$, $Cr_2O_3$, and Cr metal) and Pb-bearing compounds (i.e., $PbTiO_3$, $PbCrO_4$, $PbBr_2$, $PbCl_2$, $PbF_2$, $PbO_2$, and PbO) are also performed and used for references. The associated binding energies are obtained and listed in Table S2.

Note that there are obvious relative shifts of Cr2$p_{1/2}$ and 2$p_{3/2}$ lines toward high-energy sides from Cr metal to $Na_2CrO_4$ (Fig. S3a), indicating a gradual decrease of valence electrons because of charge screening effect. For the Cr metal with an antiferromagnetic ground state,[1] the localized 3*d* electrons should be involved and the intra-atomic Coulomb and exchange interaction are also at play to produce multiplet splitting, explaining the observed similar peak broadening to that of $Cr_2O_3$ and $CrCl_3$.

**Table S1. Determined banding energies (B.E.) of Cr2$p$ lines for the involved compounds, based on analysis of XPS data (see Fig. S3(a)).**

| Sample | Orbital | B.E. (eV) | FWHM (eV) | Orbital | B.E. (eV) | FWHM (eV) |
|---|---|---|---|---|---|---|
| Cr | $2p_{3/2}$ | 574.87 | 3.34 | $2p_{1/2}$ | 584.58 | 3.34 |
| $Cr_2O_3$ | $2p_{3/2}$ | 576.46 | 2.68 | $2p_{1/2}$ | 586.14 | 2.68 |
| $CrCl_3$ | $2p_{3/2}$ | 578.04 | 2.81 | $2p_{1/2}$ | 587.57 | 2.81 |
| $CrO_2$ | $2p_{3/2}$, $d^3\underline{L}$ | 576.71 | 2.80 | $2p_{1/2}$, $d^3\underline{L}$ | 586.19 | 2.80 |
|  | $2p_{3/2}$, $d^2$ | 579.51 | 3.00 | $2p_{1/2}$, $d^2$ | 588.92 | 3.00 |
| $Na_2CrO_4$ | $2p_{3/2}$ | 579.04 | 1.76 | $2p_{1/2}$ | 588.23 | 1.76 |
| $PbCrO_3$ | $2p_{3/2}$, $d^3\underline{L}$ | 576.20 | 2.72 | $2p_{1/2}$, $d^3\underline{L}$ | 585.78 | 2.72 |
|  | $2p_{3/2}$, $d^2$ | 578.95 | 3.00 | $2p_{1/2}$, $d^2$ | 588.10 | 3.00 |
| $Pb_{0.85}CrO_3$ | $2p_{3/2}$, $d^3\underline{L}$ | 576.43 | 2.40 | $2p_{1/2}$, $d^3\underline{L}$ | 585.89 | 2.40 |
|  | $2p_{3/2}$, $d^2$ | 578.92 | 2.11 | $2p_{1/2}$, $d^2$ | 588.14 | 2.11 |

**Table S2. Determined banding energies (B.E.) of Pb4$f$ lines for the involved compounds, based on analysis of XPS data (see Fig. S3(b)).**

| Sample | Orbital | B.E. (eV) | FWHM (eV) | Orbital | B.E. (eV) | FWHM (eV) |
|---|---|---|---|---|---|---|
| PbO | 4f$_{7/2}$, 6$p^1\underline{L}$ | 137.70 | 1.32 | 4f$_{5/2}$, 6$p^1\underline{L}$ | 142.61 | 1.32 |
|  | 4f$_{7/2}$, 6$p^0$ | 138.69 | 1.32 | 4f$_{5/2}$, 6$p^0$ | 143.59 | 1.32 |
| PbO$_2$ | 4f$_{7/2}$, 6$p^1\underline{L}$ | 137.38 | 1.00 | 4f$_{5/2}$, 6$p^1\underline{L}$ | 142.28 | 1.00 |
|  | 4f$_{7/2}$, 6$p^0$ | 138.27 | 1.70 | 4f$_{5/2}$, 6$p^0$ | 143.17 | 1.70 |
| PbF$_2$ | 4f$_{7/2}$ | 139.03 | 1.28 | 4f$_{5/2}$ | 143.93 | 1.28 |
| PbCl$_2$ | 4f$_{7/2}$, 6$p^1\underline{L}$ | 137.22 | 1.10 | 4f$_{5/2}$, 6$p^1\underline{L}$ | 142.45 | 1.10 |
|  | 4f$_{7/2}$, 6$p^0$ | 138.62 | 1.07 | 4f$_{5/2}$, 6$p^0$ | 143.52 | 1.07 |
| PbBr$_2$ | 4f$_{7/2}$, 6$p^1\underline{L}$ | 136.42 | 1.00 | 4f$_{5/2}$, 6$p^1\underline{L}$ | 141.32 | 1.00 |
|  | 4f$_{7/2}$, 6$p^0$ | 138.49 | 1.11 | 4f$_{5/2}$, 6$p^0$ | 143.39 | 1.11 |
| PbCrO$_4$ | 4f$_{7/2}$ | 138.50 | 1.10 | 4f$_{5/2}$ | 143.40 | 1.10 |
| PbTiO$_3$ | 4f$_{7/2}$, 6$p^1\underline{L}$ | 138.15 | 1.40 | 4f$_{5/2}$, 6$p^1\underline{L}$ | 143.05 | 1.40 |
|  | 4f$_{7/2}$, 6$p^0$ | 139.10 | 1.40 | 4f$_{5/2}$, 6$p^0$ | 144.05 | 1.40 |
| PbCrO$_3$ | 4f$_{7/2}$, 6$p^1\underline{L}$ | 137.50 | 1.49 | 4f$_{5/2}$, 6$p^1\underline{L}$ | 142.36 | 1.46 |
|  | 4f$_{7/2}$, 6$p^0$ | 138.66 | 1.50 | 4f$_{5/2}$, 6$p^0$ | 143.56 | 1.50 |
| Pb$_{0.85}$CrO$_3$ | 4f$_{7/2}$, 6$p^1\underline{L}$ | 137.66 | 1.88 | 4f$_{5/2}$, 6$p^1\underline{L}$ | 142.53 | 1.88 |
|  | 4f$_{7/2}$, 6$p^0$ | 138.24 | 1.83 | 4f$_{5/2}$, 6$p^0$ | 143.12 | 1.83 |

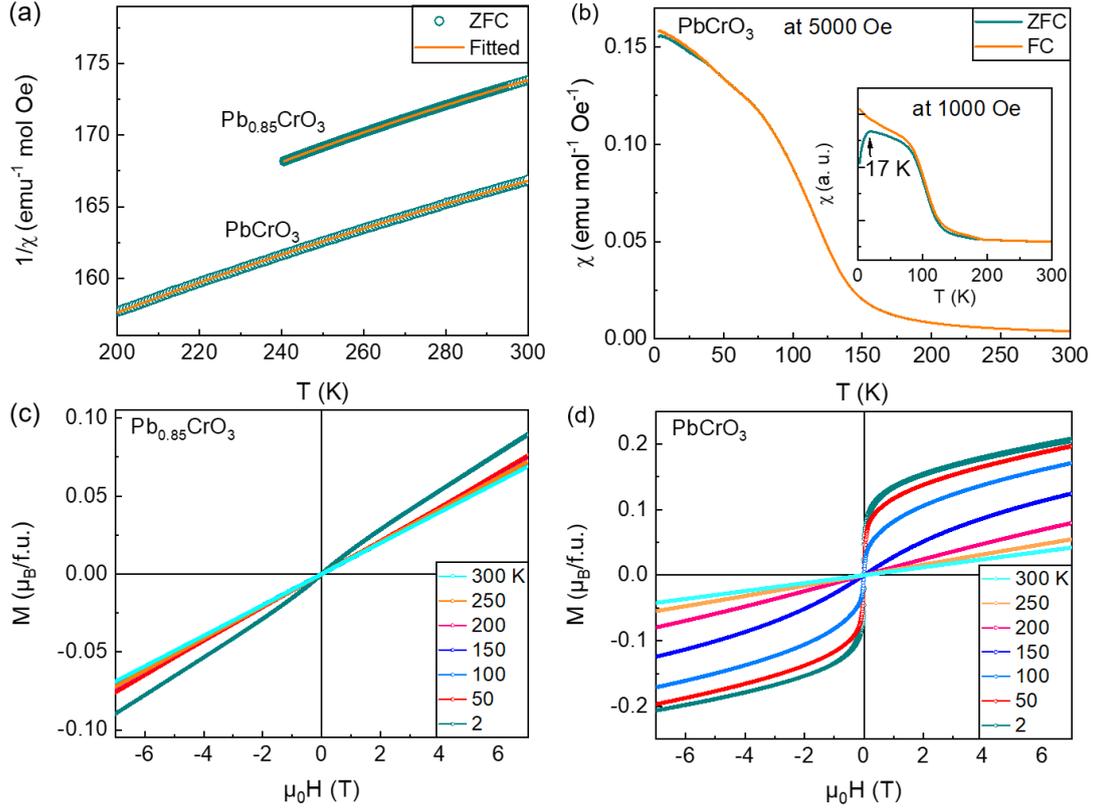

**Fig. S4. Magnetic (M) measurements of Pb$_{0.85}$CrO$_3$ and PbCrO$_3$.** (a) Inverse susceptibility (1/χ) vs. temperature at 1000 Oe. The solid lines represent fits of the selected data to the Curie-Weiss law, given by $(\chi-\chi_0)^{-1}=\frac{T-\theta_{CW}}{C}$, where variables of $\chi_0$, $\theta_{CW}$, and $C$ are the temperature-independent contribution to susceptibility, Curie-Weiss temperature, and Curie constant, respectively. (b) Magnetic susceptibility of PbCrO$_3$ taken in an external magnetic field (H) of 5000 Oe. Inset is a measurement performed at 1000 Oe, showing a possible spin re-orientation at 17 K. (c)-(d) M-H curves at different temperatures of 2, 50, 100, 150, 200, 250, and 300 K, respectively.

For each material, the effective magnetic moments, $\mu_{eff}$, can be calculated using the following equation

$$\mu_{eff}=\sqrt{\frac{3\kappa_B C}{N_A}}\mu_B,$$

where $\kappa_B$ and $N_A$ are the Boltzmann constant and Avogadro constant, respectively. Variable of $C$ is the Curie constant and can be obtained by analysis of the 1/χ-T data in Fig. S4(a). The obtained values for PbCrO$_3$ and Pb$_{0.85}$CrO$_3$ are listed in Table S3.

**Table S3. Summary of the obtained fitted magnetic constants for Pb$_{0.85}$CrO$_3$ and PbCrO$_3$ (see Fig. S4(a)).**

| Sample | $\chi_0$ | $C$ (emu K mol$^{-1}$) | $\theta_{CW}$ (K) | $\mu_{eff}$ ($\mu_B$) | Reported $\mu_{eff}$ ($\mu_B$) | Reference |
|---|---|---|---|---|---|---|
| PbCrO$_3$ | 0.00459 | 0.4859 | −117.1 | 1.97 | 2.2 | Ref. [2] |
| Pb$_{0.85}$CrO$_3$ | 0.00456 | 0.7364 | −211.7 | 2.4 | | |

The theoretically calculated $\mu_{eff}$ of Cr$^{4+}$ (i.e., 3$d^2$) is 2.82 $\mu_B$, based on the equation of $g_J\sqrt{S(S+1)}$ with $g_J = 2$ and $S = 1$.